\documentclass[final,5p,times,twocolumn]{elsarticle}

\usepackage{graphics}
\usepackage{amsmath}
\usepackage[frozencache=true,cachedir=.]{minted}
\usepackage{algorithm}
\usepackage{algpseudocode}
\usepackage{hyperref}
\hypersetup{
    colorlinks=true,
    linkcolor=blue,
}
\usepackage{braket}
\usepackage{booktabs}

\newlength{\mintednumbersep}
\AtBeginDocument{%
  \sbox0{\tiny00}%
  \setlength\mintednumbersep{\parindent}%
  \addtolength\mintednumbersep{-\wd0}%
}

\newcommand{\onlinecite}[1]{\hspace{-1 ex} \nocite{#1}\citenum{#1}}

\newcounter{bla}

\journal{Computer Physics Communications}

\begin{document}

\begin{frontmatter}



\title{PyCharge: An open-source Python package for self-consistent electrodynamics simulations of Lorentz oscillators and moving point charges}

\author{Matthew J. Filipovich\corref{author}}
\author{Stephen Hughes}

\cortext[author] {Corresponding author. \\\textit{E-mail address:} matthew.filipovich@queensu.ca}
\address{Department of Physics, Engineering Physics and Astronomy,
Queen's University, Kingston, ON K7L 3N6, Canada}

\begin{abstract}
PyCharge is a computational electrodynamics Python simulator that can calculate the electromagnetic fields and potentials generated by moving point charges and can self-consistently simulate dipoles modeled as Lorentz oscillators. To calculate the total fields and potentials along a discretized spatial grid at a specified time, PyCharge computes the retarded time of the point charges at each grid point, which are subsequently used to compute the analytical solutions to Maxwell's equations for each point charge. The Lorentz oscillators are driven by the electric field in the system and PyCharge self-consistently determines the reaction of the radiation on the dipole moment at each time step. PyCharge treats the two opposite charges in the dipole as separate point charge sources and calculates their individual contributions to the total electromagnetic fields and potentials.  The expected coupling that arises between dipoles is captured in the PyCharge simulation, and the modified radiative properties of the dipoles (radiative  decay rate and frequency shift) can be extracted using the dipole's energy at each time step throughout the simulation. 
The modified radiative properties of two dipoles separated in the near-field, which require a full dipole response to yield the correct physics, are calculated by PyCharge and shown to be in excellent agreement with the analytical Green's function results ($< 0.2\%$ relative error, over a wide range of spatial separations). Moving dipoles can also be modeled by specifying the dipole's origin position as a function of time. PyCharge includes a parallelized version of the dipole simulation method to enable the parallel execution of computationally demanding simulations on high performance computing environments to significantly improve run time. 

\end{abstract}


\end{frontmatter}

\noindent{\bf PROGRAM SUMMARY}

\begin{small}
\noindent
{\em Program Title:} PyCharge                                          \\
{\em CPC Library link to program files:} (to be added by Technical Editor) \\
{\em Developer's repository link:} \\ \href{https://github.com/MatthewFilipovich/pycharge}{github.com/MatthewFilipovich/pycharge} \\
{\em Code Ocean capsule:} (to be added by Technical Editor)\\
{\em Licensing provisions:} GPLv3 \\
{\em Programming language:}  Python 3.7 or newer         \\
{\em Supplementary material:} \\ 
Documentation is available at \href{https://pycharge.readthedocs.io}{pycharge.readthedocs.io}. The PyCharge package and its dependencies can be installed from PyPI: \href{https://pypi.org/project/pycharge/}{pypi.org/project/pycharge}   \\      
{\em Nature of problem:}\\
Calculating the electromagnetic fields and potentials generated by complex geometries of point charges, as well as the self-consistent simulation of Lorentz oscillators. \\
{\em Solution method:}\\
PyCharge calculates the individual contributions from each point charge in the system to calculate the total electromagnetic fields and potentials, and computes the dipole moment of the Lorentz oscillators at each time step by solving their governing equation of motion. \\
{\em Additional comments including restrictions and unusual features:}\\
The parallel simulation method is implemented using the mpi4py package [1]. 
   \\

\end{small}

\section{Introduction}

The majority of electrodynamics problems can be divided into two distinct classes:\footnote{There are other classes of electrodynamics problems without sources, used to obtain the underlying modes, which we are not concerned with here.} (i) one in which the goal is to solve for the electromagnetic (EM) fields generated by specified sources of charge and current (e.g., antennas, radiation from multipole sources), and (ii) one in which the motion of the charges and currents are to be determined based on the known fields in the system (e.g., motion of charges in electric and magnetic fields, energy-loss phenomena)~\cite{Jackson_1999}. However, there exists another class of electrodynamics problems where the solution requires that the fields and sources are treated self-consistently. That is, a correct treatment of the problem must include the reaction of the radiation on the motion of the sources. The self-consistent treatment of sources and fields is an old and difficult problem that stems from one of the most fundamental aspects of physics: the nature of an elementary particle. This problem of self-consistency is not only limited to classical electrodynamics, as these difficulties also arise in quantum-mechanical discussions and modelings of these systems \cite{Milonni1976}.

Motivated by the need for an electrodynamics simulator that self-consistently treats the reaction of the radiation on the real-time motion of the point charge sources, we developed the open-source Python package PyCharge. PyCharge can calculate the EM fields and potentials generated by sources in a system at specified grid points in space and time, which can then be visualized using a plotting library such as Matplotlib~\cite{mpl}. To calculate these fields and potentials, PyCharge exploits the principle of superposition in classical electrodynamics by determining the individual contributions from each source and then calculating the sum. The equations describing the scalar and vector potentials generated by a single moving point charge in a vacuum are given by the Li\'enard--Wiechert potentials, and the complete and relativistically correct equations for the time-varying electric and magnetic fields can be derived from these potentials~\cite{Griffiths_2017}.

PyCharge currently supports two types of sources: point charges that have predefined trajectories (specified as parametric equations of motion in the $x$, $y$, and $z$ directions as functions of time), and Lorentz oscillators (i.e., oscillating electric dipoles). The Lorentz oscillators (LOs) consist of two equal and opposite point charges that oscillate around the origin position (center of mass) along the axis of polarization, with a dipole moment that is dynamically calculated at each time step by solving the governing harmonic oscillator differential equation. The LOs are driven by the electric field component along the direction of polarization generated by the other sources in the system (which includes its own scattered field). As well, the LOs are naturally damped since they radiate energy as they oscillate, which dissipates kinetic energy (classically caused by {\it radiation reaction}) and decreases the dipole moment~\cite{Novotny_2006}. This damping allows PyCharge to calculate the self-consistent radiative decay rates from LOs in arbitrary motion and also in the presence of interactions with other LOs, including collective effects such as superradiance and subradiance.

The scattering of EM waves by LOs can be solved using a closed scalar and dyadic Green's function approach, where the LOs are treated as point-like objects such that their structure cannot be resolved on the scale of the wavelength of light~\cite{RevModPhys.70.447}. However, this method requires a full dipole response and cannot account for certain LO configurations (e.g., moving LOs). PyCharge simulations provide an alternative numerical method to this standard approach that yield highly accurate results and can model systems that cannot be solved analytically.
Our approach  also has notable advantages over other self-consistent EM solvers such as the finite-difference time-domain (FDTD) method~\cite{PhysRevA.95.063853}, which require a very careful treatment of the LO's divergent nature
when treated as a point dipole, which  leads to (unphysical) frequency shifts that are dependent on the grid-size.

PyCharge was designed to be accessible for a wide range of use cases: first, it can be used as a pedagogical tool for undergraduate and graduate-level EM theory courses to provide an intuitive understanding of the EM waves generated by moving point charges, and second, it can also be used by researchers in the field of nano-optics to investigate the complex interactions of light in nanoscale environments, including interactions with moving point charges and chains of resonant LOs.

We have also implemented a parallelized version of the PyCharge simulation method, using the standard Message Passing Interface (MPI) for Python package (mpi4py)~\cite{mpi_python}, which can be executed on high performance computing environments to significantly improve the run time of computationally demanding simulations (e.g., involving multiple dipoles). The PyCharge package can be installed directly from PyPI on systems running Python 3.7 or newer. Further documentation, including Python script examples and the API reference, is available at \href{https://pycharge.readthedocs.io}{pycharge.readthedocs.io}.

The rest of our paper is organized as follows: in Sec.~\ref{sec:background}, we discuss the relevant theoretical background and the applied numerical methods for calculating the EM fields and potentials generated by moving point charges; as well, we introduce the LO model for simulating dipoles and review the known effects of coupling between LOs using a photonic Green's function theory. In Sec.~\ref{sec:package_overview}, we present the general framework of the PyCharge package including the relevant classes and methods, as well as the MPI implementation. In Sec.~\ref{sec:ex_sims}, we demonstrate several electrodynamics simulations that can be performed with PyCharge and provide minimal Python listings that demonstrate PyCharge's user interface. We also verify the accuracy of simulating two coupled dipoles by comparing the calculated radiative properties and dipole energies with the known analytical solutions. Finally, we present our conclusions in Sec.~\ref{sec:conclusions}.

In addition, we provide three appendices:
\ref{APP_A} presents the  Green's function for a free-space medium
and the master equation for coupled point dipoles in a Born-Markov approximation. From these, we obtain the  key quantum electrodynamics (QED) expressions for the radiative decay rates and coupling parameters of point dipoles.
We then  provide an explicit solution to the master equation for initially excited dipoles treated as two level systems (TLSs), as these solutions demonstrate equivalence in the limit of weak excitation (linear response) with the decay dynamics of coupled LOs simulated with PyCharge.
\ref{APP_C} presents the derivation of the free-space spontaneous emission (SE) rate from the standard Fermi's golden rule approach. \ref{APP_EM_dipole} presents the exact EM fields generated by an oscillating electric dipole as functions of space and time, which we use to benchmark the accuracy of our code. 

\section{Background and methods}
\label{sec:background}
\subsection{Moving point charges}
The charge and current densities of a point charge $q$ at the position $\mathbf{r}_p(t)$ with velocity $c\boldsymbol{\beta}(t)$ are, respectively,
\begin{equation}
    \rho\left(\mathbf{r}, t\right) = q \delta\left[ \mathbf{r} - \mathbf{r}_p\right]
\end{equation}
and
\begin{equation}
    \mathbf{J}\left(\mathbf{r}, t \right) = q c\boldsymbol{\beta} \delta \left[ \mathbf{r} - \mathbf{r}_p\right],
\end{equation}
where $c$ is the vacuum speed of light.

The scalar and vector potentials of a moving point charge in the Lorenz gauge, known as the Li\'enard–Wiechert potentials~\cite{Wiechert_1901},  are derived from Maxwell's equations as
\begin{equation}\label{V}
    \Phi(\mathbf{r}, t) = \frac{q}{4\pi\epsilon_0}\left[ \frac{1}{\kappa R}\right]_{\mathrm{ret}}
\end{equation}
and
\begin{equation}\label{A}
    \mathbf{A}(\mathbf{r}, t) = \frac{\mu_0 q}{4\pi}\left[ \frac{\boldsymbol{\beta}}{\kappa R}\right]_{\mathrm{ret}},
\end{equation}
where $\epsilon_0$ and $\mu_0$ are the
vacuum permittivity and permeability, respectively,
 $R=|\mathbf{r}-\mathbf{r}_p(t')|$, and $\kappa=1-\mathbf{n}(t')\cdot \boldsymbol{\beta}(t')$ such that ${\mathbf{n}=(\mathbf{r}-\mathbf{r}_p(t'))/R}$ is a unit vector from the position of the charge to the field point, and the quantity in brackets is to be evaluated at the retarded time $t'$, given by
\begin{equation}\label{eq:ret_t}
    t' = t-\frac{R(t')}{c}.
\end{equation}

The physical (gauge-invariant) relativistically-correct, time-varying electric and magnetic fields generated by a moving point charge are, respectively,
\begin{equation}\label{E}
    \mathbf{E}\left(\mathbf{r}, t\right) = \frac{q}{4\pi\epsilon_0} \Bigg[ \frac{\left( \mathbf{n}-\boldsymbol{\beta} \right)\left(1-\beta^2\right)}{\kappa^3 R^2} + \frac{\mathbf{n}}{c\kappa^3 R} \times \left[ \left(\mathbf{n}-\boldsymbol{\beta}\right) \times \boldsymbol{\dot{\beta}} \right] \Bigg]_{\mathrm{ret}}
\end{equation}
and
\begin{equation}\label{B}
    \mathbf{B}\left(\mathbf{r}, t\right) = \frac{1}{c} \left[ \mathbf{n} \times \mathbf{E} \right]_{\mathrm{ret}},
\end{equation}
where $\boldsymbol{\dot{\beta}}$ is the derivative of $\boldsymbol{\beta}$ with respect to $t'$~\cite{Jackson_1999}. 

The first term in Eq.~\eqref{E} is known as the electric Coulomb field and is independent of acceleration, while the second term is known as the electric radiation field and is linearly dependent on $\boldsymbol{\dot{\beta}}$:
\begin{equation}\label{ECoulomb}
    \mathbf{E}_{\mathrm{Coul}}\left(\mathbf{r}, t\right) = \frac{q}{4\pi \epsilon_0}\left[\frac{\left(\mathbf{n}-\boldsymbol{\beta} \right)\left(1-\beta^2\right)}{\kappa^3 R^2}  \right]_{\mathrm{ret}}
\end{equation}
and
\begin{equation}\label{Eradiation}
    \mathbf{E}_{\mathrm{rad}}\left(\mathbf{r}, t\right) = \frac{q}{4\pi\epsilon_0c}\left[\frac{\mathbf{n}}{\kappa^3R}\times \left[ \left( \mathbf{n} - \boldsymbol{\beta} \right) \times \boldsymbol{\dot{\beta}}\right] \right]_{\mathrm{ret}}.
\end{equation}
The magnetic Coulomb and radiation field terms can be determined by substituting Eqs.~\eqref{ECoulomb} and~\eqref{Eradiation} into Eq.~\eqref{B}. Notably, the Coulomb field falls off as $1/R^{2}$, similar to the static field, while the radiation field decreases as $1/R$~\cite{Griffiths_2017}.\footnote{The conventional notation of the EM fields and potentials presented in this paper are from Ref.~\onlinecite{Jackson_1999} (Jackson); however, the PyCharge package implements these equations using the notation from Ref.~\onlinecite{Griffiths_2017} (Griffiths).}

\subsection{Computing the fields and potentials}
\label{sec:comp_fields}
PyCharge can directly calculate the EM fields and potentials generated by a moving point charge along a discretized spatial grid at a specified time. At each point on the spatial grid, the retarded time of the moving point charge, which is determined by the point charge's trajectory, is calculated using the secant method (from the SciPy package~\cite{scipy_2020}) to find the approximate solution of Eq.~\eqref{eq:ret_t}. Then, the retarded position, velocity, and acceleration of the point charge at each grid point are determined. Finally, the scalar and vector potentials are calculated from Eqs.~\eqref{V}~and~\eqref{A}, and the total, Coulomb, and radiation fields are computed using Eqs.~\eqref{E},~\eqref{ECoulomb}, and~\eqref{Eradiation} for the respective electric fields; the corresponding magnetic fields are calculated from Eq.~\eqref{B}.

In systems with multiple point charges, PyCharge exploits the superposition principle for electrodynamics simulations: the fields and potentials generated by each source are calculated using the previously described approach, and the total fields and potentials are given by the sum of the individual point charge contributions. A continuous charge density $\rho$ can be approximated in PyCharge using many point charges within the volume, where the charge value of each point charge depends on~$\rho$. Similarly, a continuous current density, described by $\mathbf{J}=\rho \mathbf{v}$, can be approximated using evenly spaced point charges traveling along a path, where the charge value of each point charge depends on $\mathbf{J}$. The accuracy of the calculated fields and potentials generated by these approximated continuous densities is dependent on both the number of point charges used in the simulation and the distance at the field point from the sources~\cite{Filipovich_2021}.

As previously discussed, PyCharge can simulate point charges that have specified trajectories defined by a parametric equation $\mathbf{r}(t) = \left(x\left(t\right), y\left(t\right), z\left(t\right)\right)$, as well as dipoles (which consist of two point charges) that are modeled as LOs with a dipole moment that is dynamically determined at each time step. In previous work~\cite{Filipovich_2021}, we simulated several interesting systems of point charges with predefined trajectories using a similar computational approach, including magnetic dipoles, oscillating and linearly accelerating point charges, synchrotron radiation, and Bremsstrahlung. The simulation of LOs in PyCharge is discussed in the next section.

\subsection{Lorentz oscillator model}
\label{sec:LO_model}
The optical interactions between light and matter at the nanometer scale are important phenomena for a variety of research fields, and a rigorous understanding of these interactions requires the use of QED theory. However, nanometer-scale structures are often too complex to be solved rigorously using only QED; in these cases, a classical approach that invokes the results of QED in a phenomenological way can be applied~\cite{Novotny_2006}. PyCharge uses the LO model, which is an approximation from quantum theory that can be derived (e.g.,  from the time-dependent Schr\"odinger equation or a quantum master equation, see~\ref{APP_A}) to simulate the interaction of a bound charge (e.g., an electron) with light~\cite{Milonni_Eberly_2010}. 

In the classical model, an oscillating dipole produces EM radiation which dissipates energy and modifies the self-consistent dipole moment. The recoil force, $\mathbf{F}_\mathrm{r}$, acting on the accelerating point charges in the dipole is called the radiation reaction or radiation damping force. The equation of motion for an undriven LO (e.g., in a vacuum) that includes the radiation reaction force is given by
\begin{equation}\label{eq:underiven_EOM}
   m\mathbf{\ddot r_{\rm dip}}(t) + \omega_0^2 m \mathbf{r_\mathrm{dip}}(t) = \mathbf{F}_\mathrm{r}(t),
\end{equation}
where $\mathbf{r}_{\rm dip}$ is the displacement from the LO's negative charge to positive charge and $\mathbf{\ddot r_{\rm dip}}$ is its second derivative with respect to time, $m$ is the effective mass of the LO (further discussed below), and $\omega_0$ is the natural angular frequency of the LO~\cite{Novotny_2006}. 

The radiation reaction force, $\mathbf{F}_\mathrm{r}$, acting on the accelerating point charges in the dipole is described by the Abraham-Lorentz formula for non-relativistic velocities:
\begin{equation}\label{eq:AL}
    \mathbf{F}_\mathrm{r}(t)=\frac{q^2}{6\pi\epsilon_oc^3}\mathbf{\dddot r_\mathrm{dip}}(t),
\end{equation}
where $\mathbf{\dddot r_\mathrm{dip}}$ is the third derivative of the displacement between the two charges~\cite{Griffiths_2017}. We can perform the approximation ${\mathbf{\dddot r}_\mathrm{dip}\approx -\omega_0^2\mathbf{\dot r}_\mathrm{dip}}$ in Eq.~\eqref{eq:AL} if the damping on the point charges introduced by the radiation reaction force is negligible (i.e., $|\mathbf{F}_\mathrm{r}| \ll \omega_0^2 m |\mathbf{r}_\mathrm{dip}|$), such that the following condition is satisfied:
\begin{equation}\label{eq:dipole_condition}
    \frac{q^2 \omega_0}{m} \ll 6\pi \epsilon_0 c^3.
\end{equation}

In an inhomogeneous environment, an oscillating electric dipole will experience the external electric field $\mathbf{E}_\mathrm{d}$ as a driving force, which is the component of the total electric field in the polarization direction at the dipole's origin (center of mass) position $\mathbf{R}$ generated by the other sources in the system and its own scattered field. If the condition in Eq.~\eqref{eq:dipole_condition} is satisfied, the equation of motion for a driven LO is
\begin{equation}\label{eq:LO_equation}
\mathbf{\ddot d}(t) +\gamma_{0} \mathbf{\dot d}(t) +\omega_{0}^{2} {\bf d}(t)= \frac{q^2}{m} \mathbf{E}_\mathrm{d}(\mathbf{R}, t),
\end{equation}
where ${\bf d}=q{\bf r_{\rm dip}}$  is the dipole moment, $\mathbf{\dot d}$ and $\mathbf{\ddot d}$ are the first and second derivatives of $\mathbf{d}$, and $\gamma_0$ is the free-space {\it energy decay rate} given by
\begin{equation}
    \gamma_0 = \frac{q^2\omega_0^2}{6\pi\epsilon_0c^3m}.
\end{equation}
This equation of motion for an LO corresponds to a Lorentzian atom model with transition frequency $\omega_0$ and linewidth $\gamma_0$ (where $\gamma_0 \ll \omega_0$), and is limited to non-relativistic velocities as it does not account for relativistic mass~\cite{Novotny_2006}. 

The effective mass $m$ (also called the reduced mass) of the dipole is given by
\begin{equation}
\label{eq:eff_mass}
    m=\frac{m_1 m_2}{m_1+m_2},
\end{equation}
where $m_1$ and $m_2$ are the masses of the two point charges in the dipole~\cite{Milonni_Eberly_2010}. These charges oscillate around the center of mass position $\mathbf{R}$, defined by
\begin{equation}
    \mathbf{R} = \frac{m_1\mathbf{r}_1+m_2\mathbf{r}_2}{m_1+m_2},
\end{equation}
where $\mathbf{r}_1$ and $\mathbf{r}_2$ are the positions of the two point charges. The point charge positions can therefore be defined in terms of the displacement between the two charges $\mathbf{r}_\mathrm{dip}$:
\begin{equation}
\label{eq:r_1}
    \mathbf{r}_1 = \mathbf{R} + \frac{m_2}{m_1+m_2}\mathbf{r}_\mathrm{dip}
\end{equation}
and
\begin{equation}
\label{eq:r_2}
    \mathbf{r}_2 = \mathbf{R} - \frac{m_1}{m_1+m_2}\mathbf{r}_\mathrm{dip}.
\end{equation}

It is also useful to discuss how the decay dynamics of LOs are related to those of a quantum TLS, in certain limits.
Specifically, 
in the limit of weak excitation (linear response), we can connect the
quantum mechanical equations of motion for a TLS to the
classical equations of motion by replacing $q^2/m$ with $q^2f/m$, where 
$f$ is the oscillator strength, defined by
\begin{equation}
\label{eq:f}
    f = \frac{2m \omega_0 d_0^2}{\hbar q^2},
\end{equation}
where $d_0=|\mathbf{d}(t=0)|$. We thus recover the usual expression for the SE rate $\gamma_{0,\mathrm{TLS}}$ from an excited TLS,
\begin{equation}
\label{eq:SE0}
    \gamma_{0,\mathrm{TLS}} = \frac{\omega_0^3 d_0^2}{3\pi\epsilon_0 \hbar c^3}.
\end{equation}

An alternative argument to relate the dipole moment with the radiative decay rate is to connect the total mean energy of the LO to the ground state energy of a quantized harmonic oscillator, so that
\begin{equation}
    \frac{m\omega_0^2 d_0^2}{q^2} = \frac{\hbar\omega_0}{2},
\end{equation}
yielding $q^2/m = 2\omega_0d_0^2/\hbar$, as expected from Eq.~\eqref{eq:f}. As well, the decay rate can be derived using a Fermi's golden rule approach (see \ref{APP_C}) from the interaction Hamiltonian
$H_\mathrm{int} = -{\bf d} \cdot \hat{\bf E}$, which leads to the following 
rate equations for the populations 
of an isolated TLS in a vacuum:\footnote{Note that we are ignoring thermal excitation processes, which is an excellent approximation for optical frequencies, since
$\hbar\omega_0 \gg k_{\rm B} T$, where $k_{\rm B}$
is the Boltzmann constant.}
\begin{equation}
\label{eq:n_e}
    \dot n_\mathrm{e}(t) = -\gamma_0 n_\mathrm{e}(t)
\end{equation}
and
\begin{equation}
\label{eq:n_g}
    \dot n_\mathrm{g}(t) = \gamma_0 n_\mathrm{e}(t),
\end{equation}
where $n_g$ and $n_e$ are the populations of the ground and excited states ($n_g+n_e=1$), respectively, and we neglect all other processes. In this picture,
$\gamma_0$ is also identical to the well known Einstein A coefficient~\cite{Milonni_Eberly_2010}. Therefore, the energy decay rate is equivalent to the population decay rate. 
We stress again that we can only make the connection between LO dynamics and populations of 
TLS states in a regime of weak excitation. 

The total energy $\mathcal{E}$ of a dipole, which is the sum of its kinetic and potential energies, is calculated by PyCharge using
\begin{equation}
\label{eq:dipoleE}
    \mathcal{E}(t) = \frac{m \omega_0^2}{2q^2} d^2(t) + \frac{m}{2q^2} \dot d^2(t),
\end{equation}
where $\dot d=|\mathbf{\dot d}|$. Since the total energy of a dipole ${\cal E}$ is proportional to $n_e$, the population of the excited state using the normalized total energy can be determined by PyCharge from
\begin{equation}
\label{eq:ne_energy}
     n_e(t)=\frac{\mathcal{E}(t)}{\max (\mathcal{E})}.
\end{equation}

\subsection{Coupled Lorentz oscillators}
\label{sec:background_coupled_LOs}
It is well known that an atom's surrounding environment modifies its radiative properties. In the classical model, the modification of the SE rate is generated by the scattering of the atomic field (as the LO is driven by the electric field at its origin position), while in QED theory the SE rate is stimulated by vacuum field fluctuations or radiation reaction, which partly depends on the ordering of the quantum field operators~\cite{Milonni1976}. Regardless, in the weak coupling regime (where the atom-field coupling constant is much less than the photon decay rate inside the cavity), the interactions can be treated perturbatively such that QED and classical theory yield the same results for the modification of the SE rate~\cite{Novotny_2006}. An exception is when the surrounding medium contains gain~\cite{PhysRevLett.127.013602}. The modification of radiative properties for two coupled LOs in close vicinity is given in \ref{APP_A} by invoking QED theory and using the dyadic Green's function for a dipole. 

The classical analogs of the superradiant and subradiant states of two coupled TLSs (where the dipoles are quantized)  occur when they are polarized along the same axis and begin either in phase (direction of the two dipole moments are equal) or out of phase (direction of the two dipole moments are reversed), respectively. PyCharge can calculate the frequency shift $\delta_{12}$ and  SE rate $\gamma^{\pm}$ of two coupled LOs in either {\it collective state} by curve fitting the discretized kinetic energy (KE) values, which are calculated by PyCharge at each time step, to the expected harmonic equation (which also connects to the master equation solutions shown in \ref{APP_A}) 
\begin{equation}
\label{eq:KE}
   \mathrm{KE}= Ae^{-(\gamma^{\pm} t)}\sin\left((\omega_0\pm\delta_{12}) t + \phi\right)^2,
\end{equation}
where $A$ and $\phi$ are constants necessary to accurately fit the function and are dependent on the initial conditions of the simulation. The curve fit should be performed using the kinetic energy values after a number of time steps have elapsed in the simulation to allow the scattered fields to propagate back to the LO's origin position. 
When the two coupled LOs are in the superradiant or subradiant states, the population of their excited state and their total energy $\mathcal{E}$ (related by Eq.~\eqref{eq:ne_energy}) are exponentially decaying functions with a decay rate of $\gamma^{+}$ or $\gamma^{-}$, respectively.

It is also useful to note that 
the total EM power radiated by an accelerating point charge in a vacuum (at non-relativistic speeds) can be calculated using the Larmor formula~\cite{Larmor}:
\begin{equation}
    P(t) = \frac{q^2 a^2(t)}{6 \pi \epsilon_0 c^3}.
\end{equation}
The power radiated by a dipole can also be calculated using the above equation by replacing $q^2 a^2$ with $|\mathbf{\ddot d}|^2$. Assuming that the dipoles begin oscillating at $t=0$~s, the radiated energy at time $t'$ can be calculated by integrating the radiated power from $t=0$~s to $t=t'$ (which can be approximated with PyCharge using a discrete integration). As well, if there are two or more dipoles in a system that interact, then each dipole will `absorb' a certain amount of energy $W_\mathrm{abs}$ radiated from the other dipoles. The total (constant) energy of a system that contains $N$ dipoles is the sum of the energy gains and losses of all the dipoles, given~by
\begin{equation}
    W_\mathrm{total} = \sum_{\mathrm{i}=1}^N  \left( \mathcal{E}_i(t')  - W_{\mathrm{abs},\,i}(t') + \int_0^{t'} P_i(t)\,dt \right),
\end{equation}
where $\mathcal{E}_i$ is the total energy (sum of the kinetic and potential energies) of the $i$th dipole in the system, defined by Eq.~\eqref{eq:dipoleE}.

\section{PyCharge package overview}
\label{sec:package_overview}

\begin{figure}[t]
    \centering
    \includegraphics[width=\linewidth]{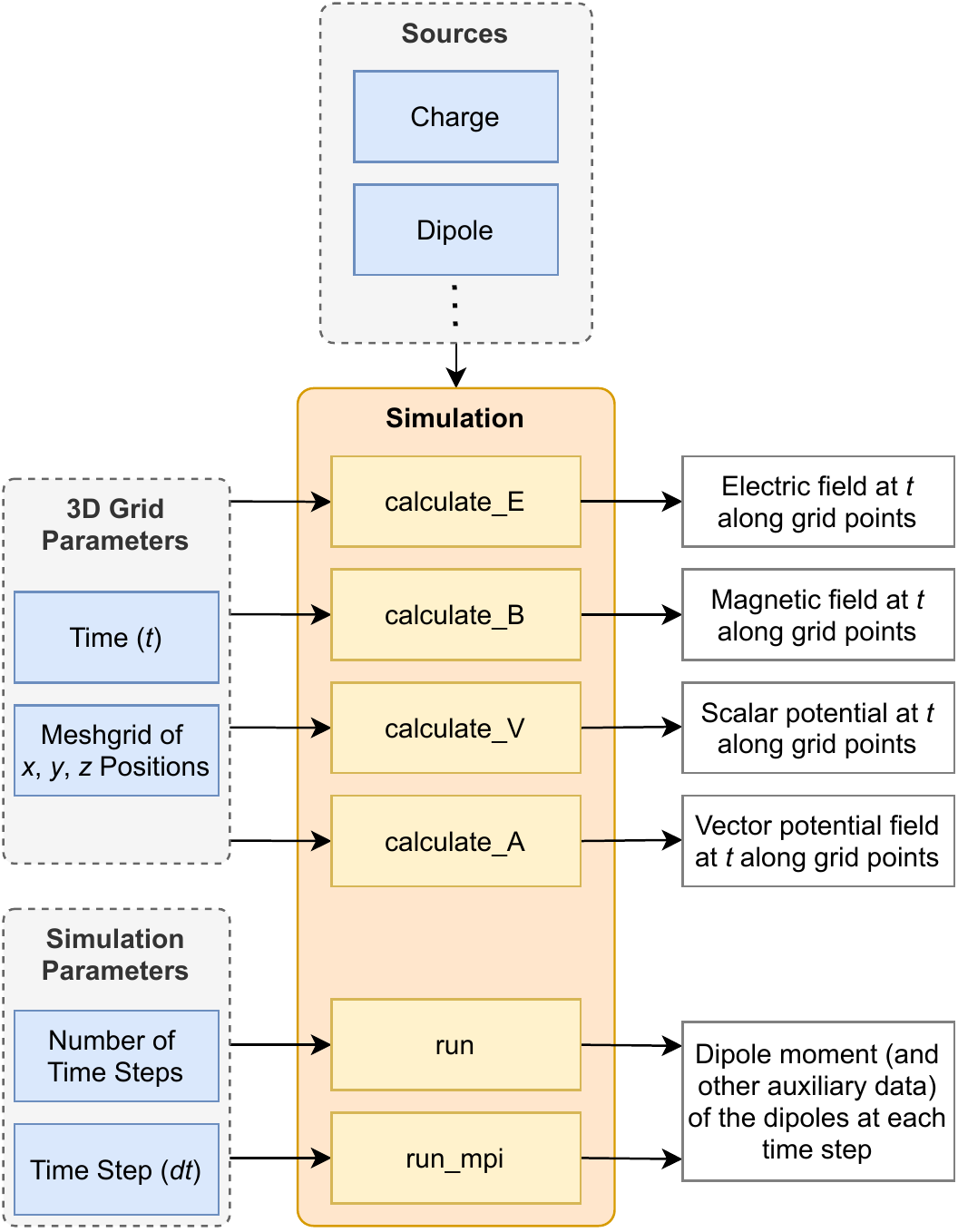}
    \caption{The \texttt{Simulation} object is instantiated with a list of the sources in the system (i.e., \texttt{Dipole} and subclasses of \texttt{Charge}). The \texttt{Simulation} object can calculate the EM fields and potentials at points along a spatial grid at a specified time $t$. The  \texttt{Simulation} object can also run (parallel) simulations to calculate the trajectory of the \texttt{Dipole} objects over a range of time steps.}
    \label{fig:workflow}
\end{figure}

PyCharge uses an object-oriented framework for representing the sources in the system and for executing simulations. All of the sources present in the system must be instantiated as objects, which include point charges with predefined trajectories and LOs (i.e., oscillating dipoles) which have dipole moments that are determined dynamically at each time step. An overview of the classes and methods implemented in the PyCharge package is shown in Fig.~\ref{fig:workflow}.

\subsection{Electromagnetic sources}

Point charge objects with predefined trajectories are instantiated from subclasses of the \verb|Charge| abstract parent class, which contains the charge $q$ as an attribute and abstract methods for the position in the $x$, $y$, and $z$ directions as functions of time. The \verb|Charge| class also has methods for the velocity and acceleration as functions of time which return the respective derivatives using finite difference approximations; however, the user can specify the exact velocity and acceleration equations in the subclasses if desired. The \verb|Charge| class also contains the method \verb|solve_time| which returns Eq.~\eqref{eq:ret_t} in a modified form and is used by PyCharge to calculate the retarded time at specified spatial points using the secant method, as discussed in Sec.~\ref{sec:comp_fields}. Several point charge classes are included with PyCharge (e.g., \verb|StationaryCharge|, \verb|OscillatingCharge|), where features of these charge trajectories (e.g., angular frequency, radius) can be modified when instantiated. Users can also create their own custom subclasses of \verb|Charge| to specify unique point charge trajectories.

The LO sources are instantiated from the \verb|Dipole| class, 
which represents a pair of oscillating point charges with a dipole moment that is dynamically determined at each time step from Eq.~\eqref{eq:LO_equation}; the positions of the point charges are then calculated using the dipole moment (Eqs.~\eqref{eq:r_1} and~\eqref{eq:r_2}). 
In PyCharge, the positive and negative charge pair are represented as \verb|_DipoleCharge| objects (which is a subclass of the \verb|Charge| class); however, they are not directly accessed by the user. The \verb|Dipole| objects are instantiated with the natural angular frequency~$\omega_0$, the origin position, the initial displacement ${\mathbf{r}}_\mathrm{dip}(t=0)$ between the two point charges in the dipole, 
the charge magnitude $q$ (default is $e=1.602 \times 10^{-19} \,$C) of the charges,
 and the mass of each charge ($m_1$ and $m_2$); the default mass for both charges is $m_e$ (with $m_e=9.109 \times 10^{-31}$~kg) such that the dipole has an effective mass of $m_e/2$ (see Eq.~\eqref{eq:eff_mass}).

 The origin position (center of mass) of the dipole can either be stationary or specified as a function of time. The \verb|Dipole| object also contains the dipole moment and its derivatives as attributes (stored as NumPy arrays), which are calculated and saved at each time step during the simulation. The dipole moment and origin position determine the motion of its two \verb|_DipoleCharge| objects, which are also updated at each time step. Unlike the point charge objects that have predefined trajectories (implemented as continuous functions), the position and related derivatives of the \verb|_DipoleCharge| objects are stored as discrete values at each time step; linear interpolation is used to calculate the values between the discrete time steps.

\subsection{Simulations}

\begin{algorithm}[b]
  \caption{\texttt{Simulation.run}}
  \label{alg:run}
   \begin{algorithmic}[1]
    \State Initialize sources in simulation
    \For{$t$ in range(0, $t_{\mathrm{max}}$, $dt$):}
        \For{dipole in sources:}
            \State Calculate $\mathbf{E}_\mathrm{d}$ and solve Eq.~\eqref{eq:LO_equation} using RK4 at $t+dt$ 
            \State Update trajectory arrays of dipole at $t+dt$
        \EndFor
    \EndFor
    \State Save Simulation and Dipole objects with trajectories
   \end{algorithmic}
\end{algorithm}

The core features of the PyCharge package, including calculating the EM fields and potentials and running simulations with \verb|Dipole| objects, are executed using the \verb|Simulation| class. The \verb|Simulation| object is instantiated with the source objects that are present in the system. The \verb|Simulation| object can calculate the electric and magnetic fields, as well as the scalar and vector potentials generated by the sources at specified spatial points at time $t$ using the methods \verb|calculate_E|, \verb|calculate_B|, \verb|calculate_V|, and \verb|calculate_A|. Additionally, the specific EM field type (Coulomb, radiation, or total field) to be calculated by the \verb|calculate_E| and \verb|calculate_B| methods can be specified. These calculations are performed using the numerical approach described in Sec.~\ref{sec:comp_fields}, and have a time complexity of $\mathcal{O}(N)$ with respect to both the number of sources in the simulation and the number of spatial points in the grid. However, the trajectories of all the sources must be defined at time $t$; therefore, the dipole moments of any \verb|Dipole| objects in the system must be known at~$t$.

\verb|Dipole| objects can be simulated in a system over a specified period of time using the \verb|run| method from the \verb|Simulation| object. The \verb|run| method calculates the dipole moment and corresponding derivatives at each time step by solving the equation of motion given in Eq.~\eqref{eq:LO_equation} using the Runge-Kutta (RK4) method. The dipoles only begin oscillating after the first time step, and have stationary dipole moments for $t\le0$ s. To calculate the driving field $\mathbf{E}_\mathrm{d}$ of each \verb|Dipole| object at a given time, the electric field generated by all of the other sources in the system is calculated at the dipole's origin position using the \verb|calculate_E| method.  Since the electric field generated by the \verb|Dipole| object must be excluded in the total field calculation to determine its own driving field $\mathbf{E}_\mathrm{d}$,  the \verb|Dipole| object is passed as a parameter to the \verb|calculate_E| method, which ensures that it does not contribute to the total field. Once the simulation is complete and the dipole trajectories are calculated at each time step, the \verb|Simulation| object and its instantiated source objects can optionally be saved using Python object serialization into an external file. The objects in the file can then be loaded by the \verb|Simulation| object for future analysis. An overview of the \verb|run| method is given in Algorithm~\ref{alg:run}.

\begin{figure}[t]
    \centering
    \includegraphics[width=\linewidth]{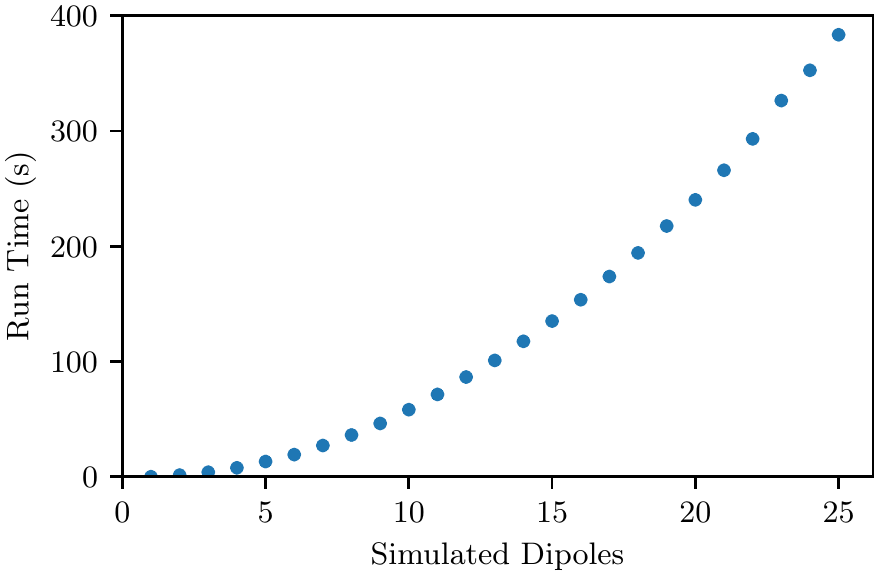}
    \caption{The average run time of the \texttt{run} method over 100 time steps with respect to the number of \texttt{Dipole} objects in the simulation. Simulations were performed using an Intel Xeon Processor E7-4800 v3 CPU.}
    \label{fig:runtime}
\end{figure}

When the \verb|run| method is called, the number of time steps and size of the time steps ($dt$) must be specified. The size of $dt$ must be appropriate for the simulation being performed: the minimum requirement is that $dt$ must be small enough such that the generated radiation does not reach the other dipoles in a single time step, and in general a smaller $dt$ value reduces the amount of error in the simulation. Other optional arguments include the name of the external file where the \verb|Simulation| object is saved after the simulation is complete (alternatively where the \verb|Simulation| object is loaded from if the simulation has already been performed), a boolean indicating whether the driving field $\mathbf{E}_\mathrm{d}$ at each time step is saved (which increases memory usage), and the maximum possible velocity achieved by the dipole's charges as the LO model does not account for relativistic effects (PyCharge raises an error if the velocity becomes larger; default is~$c/100$). The run time over 100 time steps as a function of the number of simulated \verb|Dipole| objects is shown in Fig.~\ref{fig:runtime}.

\subsection{MPI implementation}
\begin{algorithm}[b]
  \caption{\texttt{Simulation.run\_mpi}}
   \label{alg:run_mpi}
   \begin{algorithmic}[1]
    \State Initialize sources in simulation
    \State process\_dipoles = []
    \For {i in range(MPI.rank, len(dipoles), MPI.size)}
        \State process\_dipoles.append(dipoles[i])
    \EndFor
    \For{$t$ in range(0, $t_{\mathrm{max}}$, $dt$):}
        \For{dipole in process\_dipoles:}
            \State Calculate $\mathbf{E}_\mathrm{d}$ and solve Eq.~\eqref{eq:LO_equation} using RK4 at $t+dt$ 
            \State {Update trajectory arrays of dipole at $t+dt$}
        \EndFor
        \State Broadcast process\_dipoles trajectories at $t+dt$
        \State Receive and update trajectories from other dipoles
    \EndFor
    \State{Save Simulation and Dipole objects with trajectories}
   \end{algorithmic}
\end{algorithm}

Simulating the LOs using the previously described approach is {\it embarrassingly parallelizable}, as the task of solving the equation of motion (Eq.~\eqref{eq:LO_equation}) for the dipoles at each time step can be distributed across multiple processes. Ideally, each process will be tasked to calculate the trajectory of a single \verb|Dipole| object at each time step. However, if there are more \verb|Dipole| objects in the simulation than available processes, the set of \verb|Dipole| objects can be evenly distributed among the processes; in this case, the trajectories of the \verb|Dipole| objects are calculated sequentially. Once the processes have finished calculating the trajectories of their assigned \verb|Dipole| object(s), the trajectories are broadcasted to all of the other processes. The trajectories of the other dipoles, received from the other processes, are then updated for the given time step. A description of this MPI implementation is provided in Algorithm~\ref{alg:run_mpi}.

\begin{figure}[t]
    \centering
    \includegraphics[width=\linewidth]{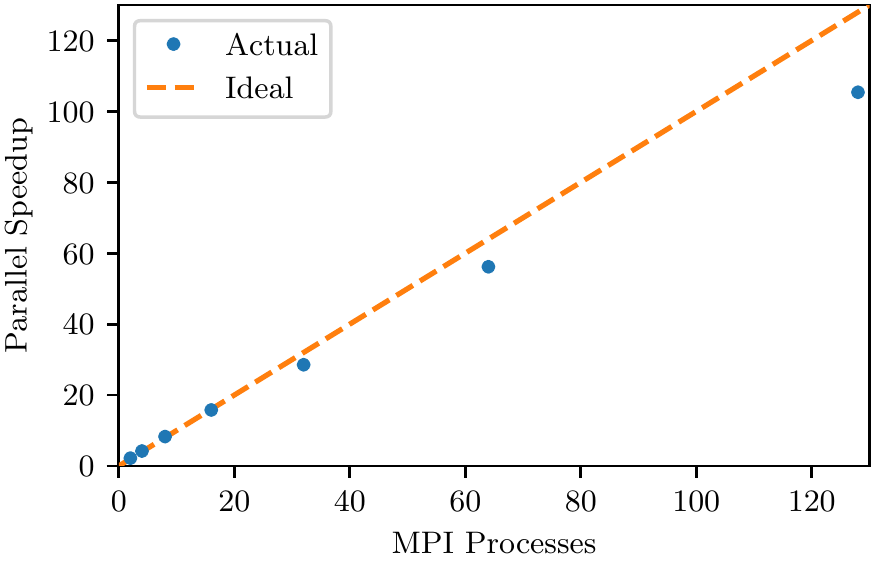}
    \caption{The average speedup of the \texttt{run\_mpi} method simulating 128 \texttt{Dipole} objects as a function of the number of MPI processes. Simulations were performed using an Intel Xeon Processor E7-4800 v3 CPU.}
    \label{fig:speedup}
\end{figure}

The original implementation of the simulation using the \verb|run| method is executed in  $\mathcal{O}(N^2)$ time for $N$ \verb|Dipole| objects, since the driving electric field $\mathbf{E}_\mathrm{d}$ of each dipole requires the calculation of the field contributions from the other $N-1$ dipoles. By taking advantage of the parallel computations, the ideal time complexity of our MPI implementation (using $N$ processes for $N$ \verb|Dipole| object) is $\mathcal{O}(N)$. However, since each process must store the trajectory arrays of the $N$ dipoles, the MPI implementation has a space complexity of $\mathcal{O}(N^2)$, while the space complexity of the original implementation is $\mathcal{O}(N)$. The average speedup offered by the MPI method using up to 128 processes is shown in Fig.~\ref{fig:speedup}. 

Future improvements to the MPI implementation could potentially reduce the space complexity to $\mathcal{O}(N)$ by pooling the dipole trajectory arrays into a single location. However, this could significantly increase the time required to fetch these trajectory values from memory.  As well, the number of broadcast operations could be reduced since it is not necessary to send the trajectory information to the other processes at each time step; instead, the trajectory values could be broadcast in batches only when they are required by the other processes, which would improve run time. 
\subsection{Performance and accuracy}

There are two main sources of numerical error in the PyCharge package: calculating the retarded time  of the sources at a given position (for determining the EM fields and potentials) by solving Eq.~\eqref{eq:ret_t} using the secant method, and determining the dipole moment at each time step for the \verb|Dipole| objects by solving Eq.~\eqref{eq:LO_equation} using the RK4 method.

The tolerance of the secant method (from the SciPy package) can be set as an initialization argument of the \verb|Simulation| object. However, the default value should be satisfactory for most simulations, typically yielding a relative error less than $10^{-6}\%$ for the fields and potentials (see Fig.~\ref{fig:osc_dipole_comp}). Extra consideration is required if the point charges are moving at relativistic velocities, as the secant method could yield a significant error. The compute time required by the \verb|calculate_E|, \verb|calculate_B|, \verb|calculate_V|, and \verb|calculate_A| methods is dependent on several factors:  as previously mentioned, the methods have a time complexity $\mathcal{O}(N)$ with respect to both the number of sources in the simulation and the number of spatial points in the grid, and also depend on the spacing of the spatial points in the grid and the trajectory of the point charges. In general, the computation time for these methods using a grid with 10$^6$~spatial points and a single \verb|Charge| object is 0.5--2~s.\footnote{Compute times recorded using an Intel Xeon Processor E7-4800 v3 CPU.\label{compute_cpu}}

The RK4 numerical method used by the \verb|run| method introduces fourth order convergence with time step size. For calculating the modified radiative properties of two coupled dipoles, we found that choosing a time step value $dt$ such that there are at least 10,000 time steps per dipole period yields relative errors less than 0.2\% (see Fig.~\ref{fig:delta_sweep}). In general, the choice of $\gamma_0$ (which is dependent on $q$, $m$, and $\omega_0$) must satisfy $\gamma_0 \ll \omega_0$ (see Eq.~\eqref{eq:dipole_condition}), and the simulation error increases with respect to the ratio $\gamma_0/\omega_0$. To accurately curve fit the kinetic energy values to the expected harmonic motion, as shown in Listing~\ref{listing:s-dipoles}, we ran simulations for four dipole periods (40,000 time steps) and used the energy values after a single dipole period (10,000 time steps) had elapsed. Using the \verb|run| method, this simulation had a run time of approximately ten minutes.\textsuperscript{\ref{compute_cpu}} Saving the simulation data into a file requires approximately 2.1~kB of memory per time step for each \verb|Dipole| object in the simulation.

\section{Example simulations}
\label{sec:ex_sims}

\begin{figure}[t]
    \centering
    \includegraphics[width=\linewidth]{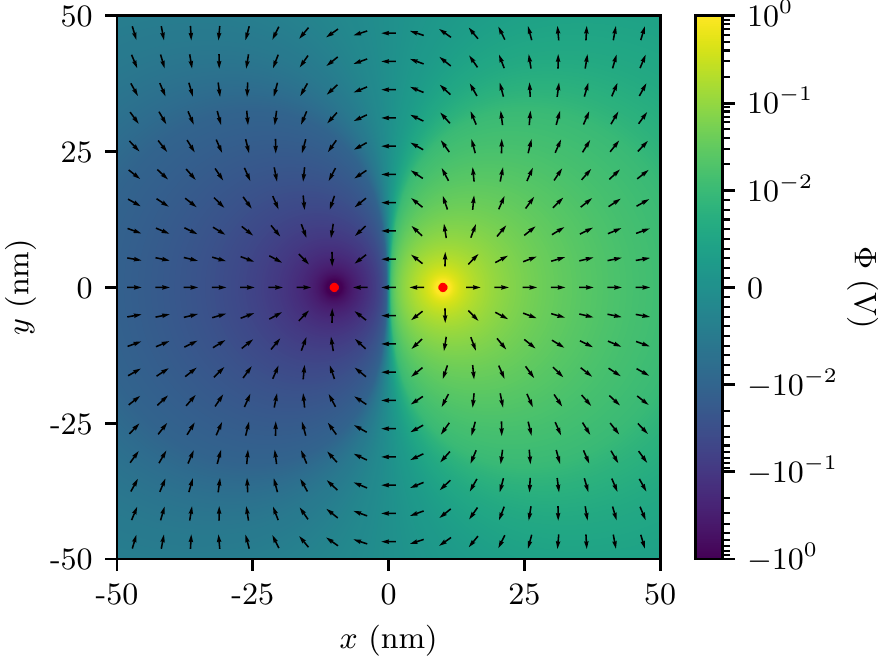}
    \caption{The scalar potential and electric field components (shown as arrows) generated by two stationary, opposite point charges (shown as red dots). The sources (with charge magnitude $e$) are separated by 20 nm along the $x$~axis. The scalar potential is plotted on a symmetrical logarithmic scale that is linear between $-10^{-2}$~V and $10^{-2}$~V.}
    \label{fig:stationary_charge}
\end{figure}

In this section, we demonstrate three different electrodynamics simulations performed using PyCharge: calculating the EM fields and potentials generated by moving point charges with predefined trajectories, simulating two coupled dipoles and determining their modified radiative properties, and instantiating moving dipoles for use in simulations. We also provide minimal Python listings that showcase the succinctness of the PyCharge interface. The Python scripts used to create the following figures can be found in the PyCharge package repository, and further examples and tutorials are available in the documentation.

\subsection{Point charges with predefined trajectories}

The EM fields and potentials generated by time-dependent point charge geometries can be complex and counterintuitive compared to their static counterparts. The calculation of the analytical solution, if one exists, often requires sophisticated vector calculus techniques that can obscure an individual's understanding and appreciation of the final result. However, using only a few lines of code, the PyCharge package allows users to calculate and visualize the full solutions to Maxwell's equations for complicated point charge geometries.

In the first example, we calculate the total electric field and scalar potential generated by two {\it stationary, opposite point charges} (i.e., a stationary electric dipole). The corresponding program code is shown in Listing~\ref{listing:stationary_charges}. The sources (two \verb|StationaryCharge| objects) are separated by 20 nm along the $x$ axis and have equal and opposite charges of magnitude $e$. The program code calculates the electric field components and scalar potential (at $t=0$ s) at each point on a $1001\times1001$ spatial grid, which is generated using the NumPy \verb|meshgrid| method. The grid is centered at the origin and extends 50 nm along the $x$ and $y$ axes. A plot of the calculated electric field components (shown as arrows) and scalar potential is shown in Fig.~\ref{fig:stationary_charge}.

\begin{figure}[t]
    \centering
    \includegraphics[width=\linewidth]{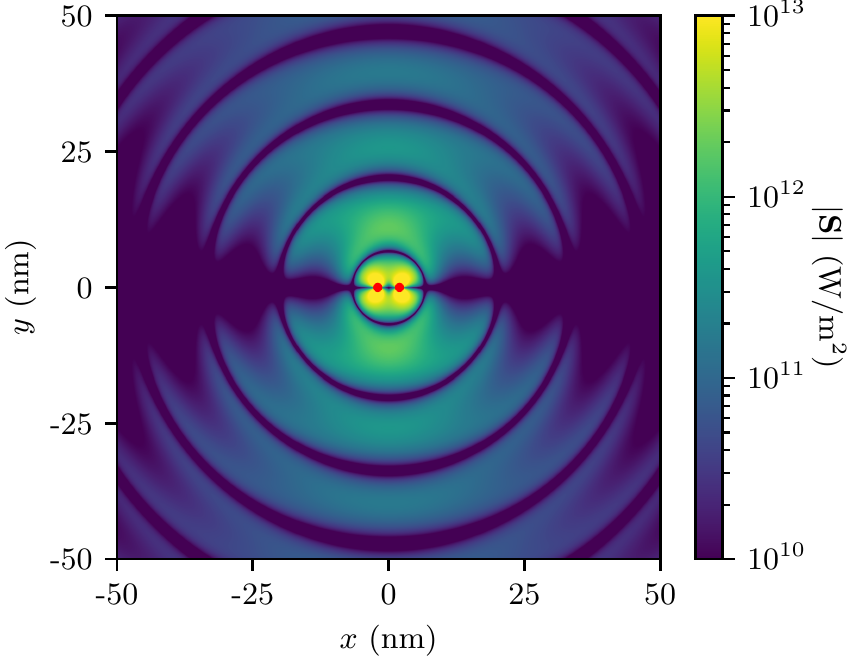}
    \caption{The magnitude of the Poynting vector of the EM fields generated by two harmonically oscillating, opposite point charges (shown as red dots). The sources (with charge magnitude $e$) oscillate around the origin with an amplitude of 2 nm and an angular frequency $\omega_0$ of $7\times10^{16}$ rad/s.}
    \label{fig:s_oscillating}
\end{figure}

\begin{listing}[H]
\begin{minted}{python}
import pycharge as pc
from numpy import linspace, meshgrid
from scipy.constants import e
sources = (pc.StationaryCharge((10e-9, 0, 0), e),
           pc.StationaryCharge((-10e-9, 0, 0), -e))
simulation = pc.Simulation(sources)
coord = linspace(-50e-9, 50e-9, 1001)
x, y, z = meshgrid(coord, coord, 0, indexing='ij')
Ex, Ey, Ez = simulation.calculate_E(0, x, y, z)
V = simulation.calculate_V(0, x, y, z)
\end{minted}
    \caption{Calculates the electric field components and scalar potential generated by two stationary point charges along a 2D spatial grid. }
    \label{listing:stationary_charges}
\end{listing}

\begin{figure}[t]
    \centering
   \includegraphics[width=\linewidth]{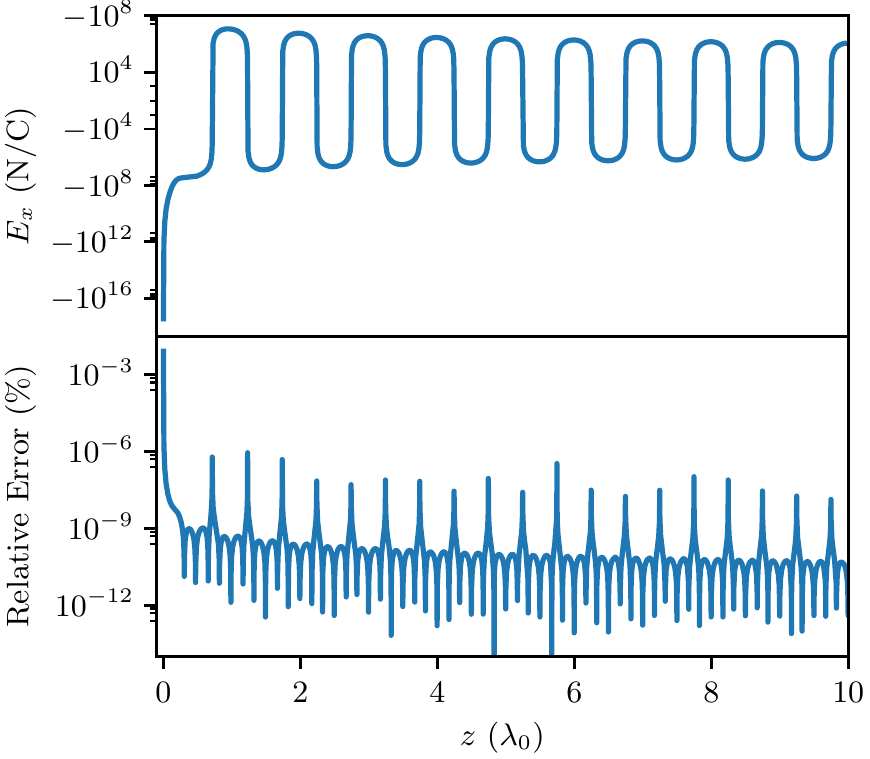}
    \caption{The $x$ component of the numerically computed electric field (top) and the respective relative error (bottom) generated by an oscillating dipole located at the origin as a function of $z$, which is scaled by the dipole's wavelength ($\lambda_0$). The theoretical values are given in Eq.~\eqref{eq:analytical_E_dipole}. The electric dipole has an angular frequency $\omega_0$ of $7\times10^{16}$ rad/s and an initial dipole moment $d_0$ of $4e\times10^{-9}$~C$\cdot$m. }
    \label{fig:osc_dipole_comp}
\end{figure}

\begin{figure}[t]
    \centering
    \includegraphics[width=0.98\linewidth]{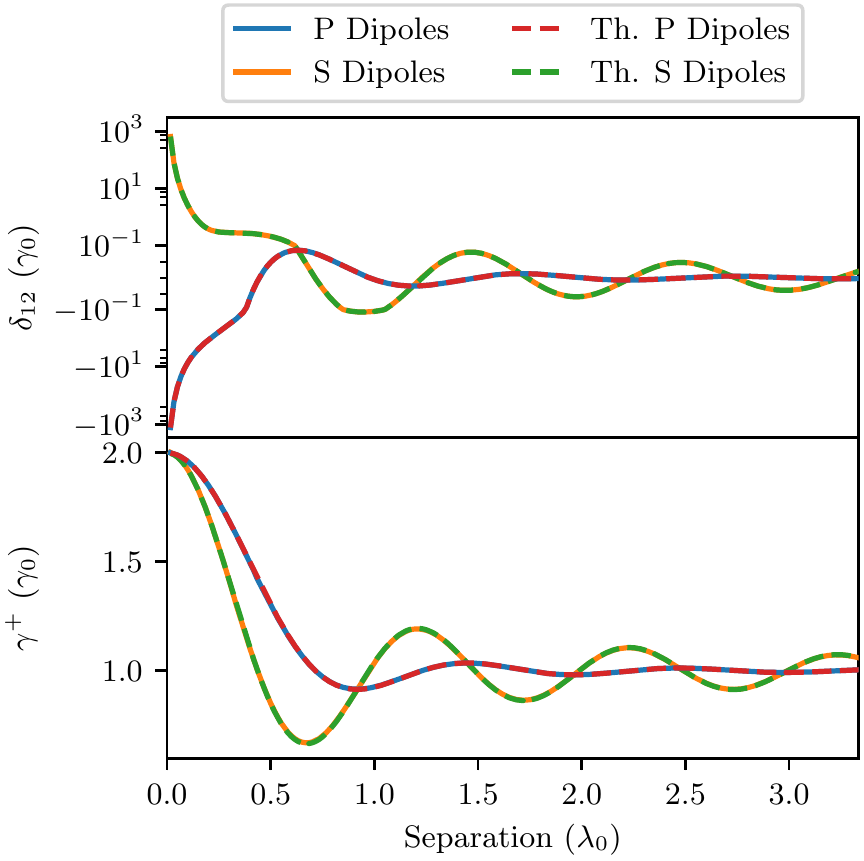}
    \caption{The simulated and theoretical frequency shift $\delta_{12}$ (top) and SE rate $\gamma^+$ (bottom) of superradiant s and p dipoles as functions of separation. The frequency shift and SE rate are scaled by the free-space decay rate $\gamma_0$, and the separation is scaled by the dipole's wavelength $\lambda_0$. The value of $\gamma_0$ for the dipoles is 19.791 MHz ($q=e$, $m=m_e/2$, and $\omega_0=200\pi\times10^{12}$ rad/s). The frequency shift is plotted on a symmetrical logarithmic scale that is linear between $-10^{-1}$~$\gamma_0$ and $10^1$~$\gamma_0$. The theoretical values for $\gamma^+$ and $\delta_{12}$ are calculated by PyCharge using Eqs.~\eqref{eq:qed_gamma_ab} and \eqref{eq:qed_delta}. The average relative errors of the $\delta_{12}$ and $\gamma^+$ values for the p dipoles are 0.15\% and 0.04\%, and for the s dipoles are 0.19\% and 0.13\%.}
    \label{fig:delta_sweep}
\end{figure}

The fields and potentials generated by different charge configurations can be simulated using the same code by instantiating other types of sources. For example, we can simulate a harmonically oscillating electric dipole by instantiating two \verb|OscillatingCharge| objects with opposite charge values ($q$) in the simulation. Users can also instantiate point charges with custom trajectories by creating a subclass of the \verb|Charge| class and defining its motion along the $x$, $y$, and $z$ directions as functions of time. 

Once the electric and magnetic fields in the system have been determined by PyCharge, we can calculate the Poynting vector $\mathbf{S}$ (the directional energy flux of the EM fields), defined by
\begin{equation}
    \mathbf{S}=\frac{1}{\mu_0}\mathbf{E}\times\mathbf{B}.
\end{equation}
The magnitude of the Poynting vector from the EM fields generated by an oscillating electric dipole  with an initial dipole moment $d_0$ of $4e\times10^{-9}$ C$\cdot$m and an angular frequency $\omega_0$ of $7\times10^{16}$~rad/s is shown in Fig.~\ref{fig:s_oscillating}. 

Additionally, the $x$ component of the electric field generated by the oscillating electric dipole along the $z$ axis, and its relative error compared to the known analytical solution (given in Eq.~\eqref{eq:analytical_E_dipole}), are shown in Fig.~\ref{fig:osc_dipole_comp}. The analytical solution describes an idealized electric dipole where the separation between the charges is infinitesimal; thus, to reduce the relative error in the near-field, the PyCharge simulation uses a separation of $4e\times10^{14}$ m and a charge value $q$ of $e\times10^5$ C, which recovers the inital dipole moment $d_0$ of $4e\times10^{-9}$ C$\cdot$m. Using these separation values, the relative error remains less than $10^{-6}\%$ but diverges in the very near-field of the dipole; this is expected since PyCharge is simulating a physical dipole with a non-infinitesimal separation.

\subsection{Two coupled dipoles}

In this section, we simulate two coupled dipoles (modeled as LOs) in a system and calculate their modified radiative properties. An example program code for simulating two s dipoles (transverse), which are polarized along the $y$ axis and separated by 80 nm along the $x$ axis, is shown in Listing~\ref{listing:s-dipoles}. The two dipoles have a natural angular frequency $\omega_0$ of $200\pi\times10^{12}$~rad/s and are simulated over 40,000 time steps (with a time step $dt$ of $10^{-18}$~s). The two charges in the dipole both have a mass of $m_e$ (the effective mass of the dipole is $m_e/2$) and a charge magnitude of $e$. Once the simulation is complete, the \verb|Simuation| and related source objects are saved into the file \verb|s_dipoles.dat|, which can be accessed for analyses. The dipoles begin oscillating in phase with an initial charge displacement $\mathbf{r}_{\mathrm{dip}}$ of 1~nm, resulting in superradiance and a modified SE rate $\gamma^+$. The rate $\gamma^+$ and frequency shift $\delta_{12}$ are then calculated in PyCharge by curve fitting the kinetic energy of the dipole (using the kinetic energy values after the 10,000 time step), as discussed in Sec.~\ref{sec:background_coupled_LOs}. As well, the theoretical values for $\gamma_{12}$ (related to $\gamma^+$ by Eq.~\eqref{eq:gamma_super_sub}) and $\delta_{12}$ are calculated by PyCharge using Eqs.~\eqref{eq:qed_gamma_ab} and \eqref{eq:qed_delta}.

\begin{figure}[t]
    \centering
    \includegraphics[width=\linewidth]{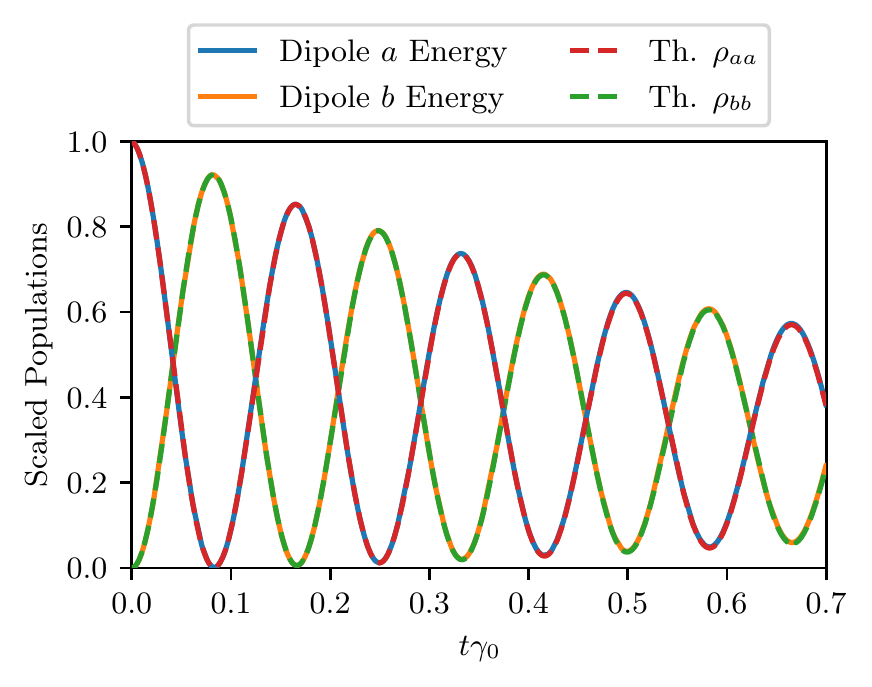}
    \caption{The normalized populations of the excited states of two dipoles $a$ and $b$, where dipole $a$ is initially excited ($\rho_{aa}(0)=1$) and dipole $b$ is not (${\rho_{bb}(0)=0}$). The dipoles are separated by 80 nm (0.053 $\lambda_0$) and have a natural angular frequency $\omega_0$ of $400\pi\times10^{12}$ rad/s. The free-space decay rate $\gamma_0$ of the dipoles is 7.916 GHz ($q=20e$ and $m=m_e/2$). The total energy is calculated using Eq.~\eqref{eq:dipoleE}, and the analytical solutions for the excited state populations are given in Eqs.~\eqref{eq:rho_aa} and~\eqref{eq:rho_bb}.}
    \label{fig:ca_cb}
\end{figure}

\begin{figure}
    \centering
    \includegraphics{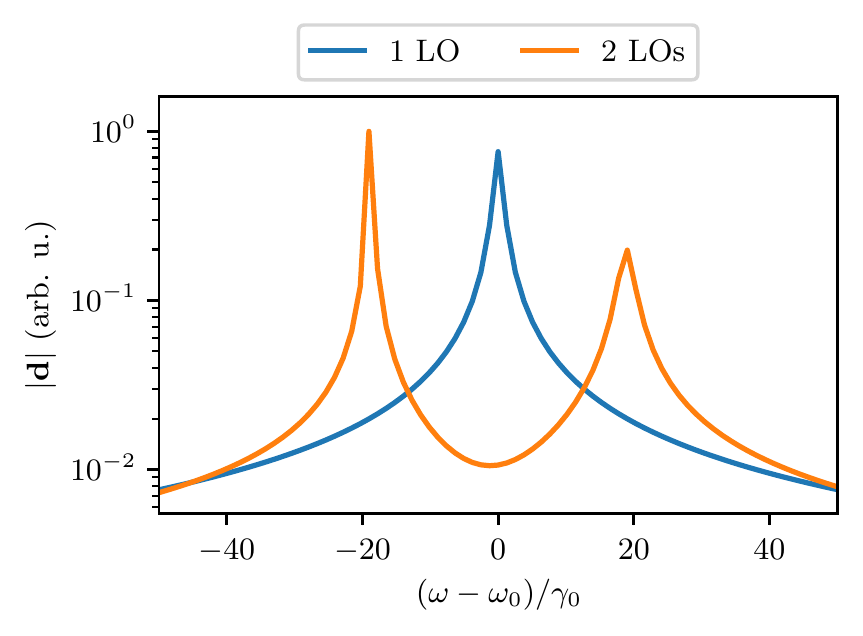}
    \caption{The dipole moment in the frequency domain for one isolated LO (free-space decay) and two coupled LOs in free-space, where the latter response clearly shows the subradiant (lower energy resonance) and supperradiant states (higher energy resonance). The two LOs are separated by 80 nm and both have angular frequencies of $400\pi\times10^{12}$ rad/s, and the theoretical (scaled) frequency shift $\delta_{12}$ is 18.86~$\gamma_0$.
     }
    \label{fig:fft}
\end{figure}

The radiative properties of two coupled dipoles as a function of separation can be calculated by repeatedly running the previous simulation while sweeping across a range of dipole separation values. Using this technique, the modified rate $\gamma^+$ and frequency shift $\delta_{12}$ for in phase (superradiant) s and p dipoles, scaled by the free-space emission rate $\gamma_0$, are plotted in Fig~\ref{fig:delta_sweep}. The theoretical results from QED theory are also shown in the figure, and the relative errors values ($< 0.2\%$) are provided.

\begin{listing}[H]
\begin{minted}{python}
import pycharge as pc
from numpy import pi
timesteps = 40000
dt = 1e-18
omega_0 = 100e12*2*pi
origins = ((0, 0, 0), (80e-9, 0, 0))
init_r = (0, 1e-9, 0)
sources = (pc.Dipole(omega_0, origins[0], init_r),
           pc.Dipole(omega_0, origins[1], init_r))
simulation = pc.Simulation(sources)
simulation.run(timesteps, dt, 's_dipoles.dat')
d_12, g_plus = pc.calculate_dipole_properties(
    sources[0], first_index=10000)
d_12_th, g_12_th = pc.s_dipole_theory(
    r=1e-9, d_12=80e-9, omega_0=omega_0)
\end{minted}
    \caption{Runs the simulation of two coupled (in phase)  s~dipoles and calculates their radiative properties, as well as the theoretical radiative results from QED theory. From the code: $\delta_{12}=156.919$, $\delta_{12,\mathrm{th}}=156.926$, $\gamma^+=1.997$, and $\gamma_{12, \mathrm{th}}=0.994$ (scaled in units of $\gamma_0$).}
    \label{listing:s-dipoles}
\end{listing}

We can also plot the normalized populations of the excited states of two coupled dipoles, $\rho_{aa}(t)$ and $\rho_{bb}(t)$, using the normalized total energy of the dipoles at each time step (Eqs.~\eqref{eq:dipoleE} and~\eqref{eq:ne_energy}). This yields particularly interesting results for coupled dipoles with small separations when one dipole is initially excited ($\rho_{aa}(0)=1$) and the other is not ($\rho_{bb}(0)=0$). In this scenario, the populations are a linear combination of the superradiant and subraddiant states, which leads to the observed energy transfer between dipoles known as F\"orster coupling,\footnote{The solution calculated by PyCharge is more general as we also include dynamical coupling terms beyond the usual $1/|r|^3$ static coupling regime, but the F\"orster coupling is fully recovered.  Indeed for chains of coupled dipoles, the retardation effects become essential to include~\cite{Citrin1995}.} as further discussed in~\ref{APP_A}. This phenomenon can be simulated in PyCharge by initializing the excited dipole with a much larger dipole moment (and total energy) than the other. The simulation results and analytical solution, given in Eqs.~\eqref{eq:rho_aa} and~\eqref{eq:rho_bb}, are shown in Fig.~\ref{fig:ca_cb}. Additionally, the dipole moment of dipole $a$  in the frequency domain is shown in Fig.~\ref{fig:fft}, which clearly shows the frequency peaks of the subradiant and supperradiant states.\footnote{An identical frequency plot could also be created using the dipole moment of dipole $b$.} The dipole moment of an isolated LO in the frequency domain is also shown for comparison.

\subsection{Moving dipoles}
In addition to stationary dipoles, PyCharge can self-consistently simulate moving dipoles (e.g., oscillating) with a time-dependent origin (center of mass) position. 
Other direct EM simulation approaches (e.g., the FDTD method) cannot accurately model moving dipoles, which can have practical importance for nano-scale interactions as real atoms are rarely stationary. Thus, PyCharge can be used to explore new physics phenomena that arise from this additional dipole motion (e.g., phonons in dipole chains).
Simulations with moving dipoles are performed in PyCharge by creating a function that accepts the time $t$ as a parameter and returns the position of the dipole's origin position at $t$ as a three element array ($x$, $y$, $z$). This function is then passed as a parameter when instantiating the \verb|Dipole| object. An example of instantiating a \verb|Dipole| object with a time-dependent origin is given in Listing~\ref{listing:moving_dipole}. A detailed analysis of moving dipoles using the PyCharge package will appear in future work.

\begin{listing}[H]
\begin{minted}{python}
from numpy import pi, cos
import pycharge as pc
def fun_origin(t):
    x = 1e-10*cos(1e12*2*pi*t)
    return ((x, 0, 0))
omega_0 = 100e12*2*pi
init_d = (0, 1e-9, 0)
source = pc.Dipole(omega_0, fun_origin, init_d)
\end{minted}
    \caption{Instantiates a \texttt{Dipole} object with a time-dependent origin position that oscillates along the $x$ axis with an amplitude of 0.1~nm and an angular frequency of $2\pi\times10^{12}$~rad/s.}
    \label{listing:moving_dipole}
\end{listing}

\section{Conclusions}
\label{sec:conclusions}
PyCharge was developed as an open-source simulation package to allow both novice and experienced users model a wide range of classical electrodynamics systems using point charges. PyCharge can calculate the time-dependent, relativistically correct EM fields and potentials generated by moving point charges with predefined trajectories. The user can create custom point charge objects in PyCharge by defining the $x$, $y$, and $z$ charge positions as functions of time. PyCharge can also self-consistently simulate the motion of LOs (dipoles), which are driven by the electric field generated by the other sources in the system. With only a few lines of code to set up the simulation, PyCharge can return the calculated modified radiative properties of the LOs (SE rates and frequency shift) in the system.

Simulating multiple LOs in PyCharge is numerically exact and does not rely on a Markov approximation, which has clear advantages for scaling to multiple dipoles where analytically solving chains of atoms via coupling rates and master equations becomes tedious and eventually intractable. As well, the origin position of the LOs can be stationary or time-dependent, and the latter is often very difficult to calculate analytically. We hope that PyCharge will prove useful as a novel simulator in the rapidly advancing field of computational electrodynamics, and expect that future versions of PyCharge will be improved by implementing new ideas from the open-source research community.

\section*{Acknowledgements}
This work was supported by the Natural Sciences and Engineering Research Council of Canada, Queen's University, and the Canadian Foundation for Innovation. We also acknowledge support from CMC Microsystems, Xanadu Quantum Technologies, and Mitacs, as well as from the Centre for Advanced Computing (CAC) at Queen's University.

\appendix
\section{Green's functions, quantum master equations, and   analytical expressions for the radiative decay rates and coupling parameters}
\label{APP_A}

\subsection{Green's function for free-space}
To describe the general theory of light emission, we first define the dyadic Green's function $\mathbf{G}(\mathbf{r}, \mathbf{r'}; \omega)$,  which describes the field response at $\mathbf{r}$ to an oscillating polarization dipole at $\mathbf{r'}$ as a function of frequency. The Green's function is the solution to the wave equation
\cite{RevModPhys.70.447,Yao_Hughes_2010,Vlack2012}
\begin{equation}
\left[\nabla \times \nabla \times-\frac{\omega^{2}}{c^{2}} \epsilon(\mathbf{r})\right] \mathbf{G}\left(\mathbf{r}, \mathbf{r}^{\prime}, \omega\right)=\frac{\omega^{2}}{c^{2}} \mathbf{I} \delta\left(\mathbf{r}-\mathbf{r}^{\prime}\right),
\end{equation}
where $\mathbf{I}$ is the unit dyadic , and $\epsilon=n^2$ is the dielectric constant that we will assume is lossless (real), and we also assume a non-magnetic material. For a homogeneous dielectric with a refractive index $n$ (where $n=1$ in a free-space medium), the homogeneous Green's function can be written analytically given the wavevector in the background medium $k=\omega n/c$:
\begin{equation}
\begin{aligned}
{\mathbf{G}}_{\mathrm{hom}}(R ; \omega) & = \left({\mathbf{I}}+\frac{\nabla \nabla}{k^{2}}\right) \frac{k_{0}^{2} e^{i k R}}{4 \pi R} \\
& = \frac{{\mu_{0} k_{0}^{2}} \exp \left(i k R\right)}{4 \pi R}\left[\left(1+\frac{i k R-1}{k^{2} R^{2}}\right) {\mathbf{I}} \right. \\
&+ \left.\left(\frac{3-3 i k R-k^{2} R^{2}}{k^{2} R^{2}}\right) \frac{\mathbf{R} \otimes \mathbf{R}}{R^{2}}\right] + \frac{\delta(R)}{3 n^2}{\mathbf{I}}  ,
\end{aligned}
\end{equation}
where $R=|\mathbf{R}|=|\mathbf{r}-\mathbf{r'}|$ and $k_0=\omega/c$.

Although it is possible to analytically define the exact time-dependent solution
from a Fourier transform of 
an exact Dyson solution in the presence of 
a finite number of quantum emitters (treated as quantized harmonic oscillators), and thus obtain an exact solution to the emitted spectrum~\cite{Wubs_2004,Yao2009},
below we present a simpler and more common solution (by invoking a Markov approximation) that immediately connects to the main physics regimes studied in this paper.

\subsection{Quantum master equation for coupled two level systems and the coupling rates}

In QED, treating the atoms as TLSs, one can use a Born-Markov approximation to derive the master equation for the reduced density $\rho$, where the decay rates $\gamma_{ij}$ appear directly as Lindblad superoperators, and $\delta_{12}$ is a simple frequency shift
$\omega_i \rightarrow \omega_i + \delta_{ij}$.
For two coupled
TLSs, $a$ and $b$, the resulting master equation
(in the interaction picture) is~\cite{PhysRevA.12.1475,PhysRevA.91.051803,Gangaraj2015}
\begin{align}
    \frac{d \rho}{dt}
    &= \sum_{\alpha,\,\beta=a,\,b}
    \frac{\gamma_{\alpha\beta}(\omega_\alpha)}{2}
    \left [2 \sigma^-_\alpha \rho \sigma^+_\beta - \sigma^+_\alpha\sigma^-_\beta \rho - \rho\sigma^+_\alpha \sigma^-_\beta \right] \nonumber \\
    & - i 
     \left [ \left (\delta_{ab}(\omega_b) \sigma^+_a \sigma^{-}_b
    + \delta_{ba}(\omega_a) \sigma^+_b \sigma^{-}_a \right ),\rho \right],
\end{align}
where $\sigma^\pm_{\alpha}$ and $\sigma^\pm_{\beta}$ are the Pauli operators for the TLSs
(i.e., $\sigma^+_\alpha=\ket{e_\alpha}\bra{g_\alpha}$ and $\sigma^-_\alpha=\ket{g_\alpha}\bra{e_\alpha}$).
The master equation accounts for the interactions between the quantum emitters
and the surrounding environment, and we have also used a rotating wave approximation. 

For two coupled TLSs (which recover the same model as two quantized harmonic oscillators in the weak excitation
approximation), $a$ and $b$, in close vicinity with equal resonance frequencies ($\omega_0 = \omega_a= \omega_b$), the self ($\gamma_{aa,bb}$) and cross ($\gamma_{ab,ba}$) decay rates  are obtained from~\cite{PhysRevA.12.1475,PhysRevA.91.051803, PhysRevA.66.063810}
\begin{align}
\label{eq:qed_gamma_ab}
    \gamma_{\alpha \beta} &=  \frac{2{\bf d}_\alpha^\dagger  \cdot {\rm Im} {\bf G}({\bf r}_\alpha,{\bf r}_\beta,\omega_0)  \cdot {\bf d}_\beta}{\epsilon_0 \hbar}.
\end{align}

Assuming the two TLSs are identical ($\mathbf{d}_a=\mathbf{d}_b$ and $\omega_a=\omega_b$), we define the on-resonance Markovian decay rates as
$\gamma_{0}\equiv\gamma_{aa}= \gamma_{bb}$ and $\gamma_{12}\equiv\gamma_{ab}=\gamma_{ba}$. The hybrid system (in the presence of coupling) can then form
superradiant or subradiant states~\cite{PhysRev.93.99}, defined from $\ket{\psi^+}=1/\sqrt{2}\,(\ket{e_a,g_b}
+ \ket{g_a,e_b})$ and $\ket{\psi^-}=1/\sqrt{2}\,(\ket{e_a,g_b}
- \ket{g_a,e_b})$, respectively, which decay with the modified rates
\begin{equation}
\label{eq:gamma_super_sub}
\gamma^{\pm}
= \gamma_{0} \pm \gamma_{12}.
\end{equation}

The so-called virtual photon transfer rate (or dipole-dipole induced frequency shift) between two TLSs with equal resonance frequencies is
\begin{equation}
\label{eq:qed_delta}
    \delta_{\alpha\beta}|_{\alpha \neq \beta}
    = -\frac{{\bf d}_\alpha^\dagger \cdot {\rm Re} {\bf G}({\bf r}_\alpha, {\bf r}_\beta,\omega_0) \cdot {\bf d}_\beta}{\epsilon_0 \hbar}
    .
\end{equation}
This ``exchange'' term  fully recovers F\"orster coupling and can yield superradiant and subradiant states (Dicke states) for two coupled TLSs at small separation distances~\cite{PhysRev.93.99}.


Although the expressions in terms of the photonic Green's function are general for any medium, to recover the free-space dipole problem in the main text, we simply replace 
${\bf G}$ by ${\bf G}_{\rm hom}$ and obtain these rates analytically (within a Markov approximation, i.e.,  evaluated at a single frequency).
We can rewrite the quantum master equation for two coupled TLSs with equal resonance frequencies as
\begin{align}
\label{eq:ME}
    \frac{d \rho}{dt}
    =&\sum_{\alpha,\,\beta=a,\,b}
    \frac{\gamma_{\alpha\beta}}{2}
    \left [2 \sigma^+_\alpha \rho \sigma^-_\beta - \sigma^+_\alpha\sigma^-_\beta \rho - \rho\sigma^+_\alpha \sigma^-_\beta \right] \nonumber \\
    & - i \delta_{12}
    \left [ \left (\sigma^+_a \sigma^{-}_b
    +  \sigma^+_b \sigma^{-}_a \right ),\rho \right].
\end{align}

From the master equation,
we can easily derive the equation 
of motion for any observable of interest, i.e.,
from $\braket{\dot O}= \braket{\dot \rho  O} = {\rm Tr}(\dot \rho  O)$.
For example, the population equation of motion for the two coupled dipoles are
\begin{align}
\label{eq:DMpops}
    \dot \rho_{aa}
    & = -\gamma_{aa}\rho_{aa}-\gamma_{ab}
    \rho_{ab} - i \delta_{ab} \rho_{ab},  \\
    \dot \rho_{bb}
    &= -\gamma_{bb}\rho_{bb}-\gamma_{ba}
    \rho_{ba} + i \delta_{ba} \rho_{ba},
\end{align}
where the density matrix elements are
$\rho_{\alpha\beta}= \braket{\alpha|\rho|\beta}$. 


The coherence between the TLSs, accounted for by the terms $\rho_{ab}$ and $\rho_{ba}$ (whose equations can be derived similarly), can significant affect the radiative decay rates, allowing various collective solutions such as superradiant and subradiant decays. For example, given the initial conditions $\rho_{aa}(0)=1$ and $\rho_{bb}(0)=\rho_{ab}(0)=\rho_{ba}(0)=0$, and assuming the dipoles are identical, the excited state populations
have a non-trivial time dependence with oscillatory dynamics, as shown
in Eqs.~\eqref{eq:rho_aa} and~\eqref{eq:rho_bb}.

In the next section, we will solve the density matrix equations in a different basis (using the dressed states), which both simplifies their solution and clearly shows the collective modified decay rates for the superradiant and subradiant states -- which decay with the 
rates $\gamma^+$ and $\gamma^-$, respectively.

\subsection{Time-dependent solution to the master equation for initially excited atoms}

With no initial driving field included, the reduced master equation (Eq.~\eqref{eq:ME}) can be solved analytically. To make this clear, we can restrict the size of the basis to include the following four states:
$\ket{I}= \ket{g_a,g_b}$,
$\ket{II}=\ket{e_a,e_b}$, and 
$\ket{\pm} = 1/\sqrt{2}\,(\ket{e_a,g_b}\pm\ket{g_a,e_b})$, where $g$ and $e$  label the ground and excited states of the TLSs.
If the initial excitation only involves the density matrix elements
$\rho_{++}$, $\rho_{--}$, $\rho_{+-}$,
and $\rho_{-+}$ (so only the atoms are excited, i.e., the fields are in a vacuum state, $\ket{\phi}_{\rm fields} =\ket{\{0}\}$),
with $\rho_{\alpha\beta}=\ket{\alpha}\bra{\beta}$,
then we have the following density matrix equations for two identical TLSs: 
\begin{equation}
\begin{aligned}
\label{eq:DM}
    \dot \rho_{++} &=
    -(\gamma_{0}+\gamma_{12})\rho_{++},
     \\ 
        \dot \rho_{--} &= -(\gamma_0-\gamma_{12}) \rho_{--},  \\
        \dot \rho_{+-} &=
        -(\gamma_0 + i2\delta_{12})\rho_{+-} ,   \\
       \dot \rho_{-+} &= -(\gamma_0 - i2\delta_{12})\rho_{-+}, 
\end{aligned}
\end{equation}
which have the explicit solutions 
\begin{equation}
\begin{aligned}
\label{eq:DM2}
    \rho_{++}(t) &= \rho_{++}(0) e^{-(\gamma_{0}+\gamma_{12})t},  \\ 
        \rho_{--}(t) &= \rho_{--}(0) e^{-(\gamma_{0}-\gamma_{12})t},  \\
        \rho_{+-}(t) &= \rho_{+-}(0) e^{-(\gamma_{0}+ 2i \delta_{12})t},  \\
       \rho_{-+}(t) &= \rho_{-+}(0)e^{-(\gamma_{0}- 2i \delta_{12})t}, 
\end{aligned}
\end{equation}
which is a particular case of weak excitation, so the 
two quantum state ($\ket{II}$) is decoupled. Consequently,
this coupled TLS solution recovers the solution of  coupled quantized harmonic oscillators, and  this is also why the radiative decay of classical LOs are
then identical in this limit.

These decay solutions are precisely the cases of superradiant decay, subradiant decay, and a linear combination of superradiant and subradiant decay. The latter case will cause population beatings that oscillate with a beating time  of $T_{\rm beat}=  \pi/\delta_{12}$.
Although we have derived these equations in a Markov approximation, we note that this is not necessary in general, and the full time-dependent quantum dynamics can also be worked out analytically in a weak excitation approximation~\cite{Yao2009}.
The PyCharge simulations are also numerically exact and do not rely on a Markov approximation, and have clear advantages for scaling to multiple dipoles, where analytically solving chains of atoms via coupling rates and master equations becomes tedious and eventually intractable.

The expectation values for observables 
in the original basis
are derived in the usual way, e.g., for the excited population in the TLS $a$,
we have 
\begin{equation}
   \rho_{aa}=\braket{\sigma^+_a\sigma^-_a}
    = \sum_{i,j}\bra{j}\sigma^+_a\sigma^-_a\ket{i} \rho_{ji},
\end{equation}
where $i,j$ sums over states $\ket{I},\ket{II},$ and $\ket{\pm}$,
and similarly for  $\rho_{bb}$. For an initial condition
of $\rho_{aa}(0)=1$ and $\rho_{bb}(0)=\rho_{ba}(0)=\rho_{ab}(0)=0$, this is equivalent to
having $\rho_{++}(0) =\rho_{+-}(0)+\rho_{-+}(0)=\rho_{--}(0)=1/4$. The explicit time-dependent solutions for the population decays, from Eq.~\eqref{eq:DM2}, is
\begin{equation}
\label{eq:rho_aa}
    \rho_{aa}(t) = \frac{1}{4}
    \left (e^{-(\gamma_0-\gamma_{12}) t}+e^{-(\gamma_0+\gamma_{12}) t} +  2\cos(2\delta_{12} t) e^{-\gamma_0 t} \right)
\end{equation}
and 
\begin{equation}
\label{eq:rho_bb}
    \rho_{bb}(t) = \frac{1}{4}
    \left (e^{-(\gamma_0-\gamma_{12}) t}+e^{-(\gamma_0+\gamma_{12}) t} -  2\cos(2\delta_{12} t) e^{-\gamma_0 t} \right).
\end{equation}
Finally, in the limit of very small dipole separations, where $\gamma_{12} \approx \gamma_0$,  we have the approximate solutions
\begin{equation}
    \rho_{aa}(t)  \approx \frac{1}{4}
    \left (1+e^{-2\gamma_0 t} +  2\cos(2\delta_{12} t) e^{-\gamma_0 t} \right)
\end{equation}
and
\begin{equation}
    \rho_{b}(t)  \approx \frac{1}{4}
    \left (1+e^{-2\gamma_0 t} -  2\cos(2\delta_{12} t) e^{-\gamma_0 t} \right).
\end{equation}

\section{Fermi's Golden Rule for the Free-Space Spontaneous Emission Rate}
\label{APP_C}

Here, we briefly show the
standard Fermi's golden rule approach for calculating the free-space SE rate.
Fermi's golden rule is written as
\begin{equation}
    \gamma_{i \rightarrow f}(\omega_f)
    =\frac{2\pi}{\hbar}
    \left |\braket{i|H_{\rm int}|f} \right |^2 D(\omega_f),
\end{equation}
where $D$ is the density of states (assumed to be approximately constant over the region of emission), and $i$
and $f$ are the initial and final states, respectively. Consistent with the
Markov approximation in the density matrix approach, this is also a long time Markovian ``rate''.

The dipole interaction
Hamiltonian $H_{\rm int}$ has the usual form
\begin{align}
    H_{\rm int} = -\sum_{{\bf k},\eta} \sqrt{\frac{\hbar\omega_{\bf k}}{2 \epsilon_0}}
   \left(\sigma^++\sigma^- \right)
    {\bf d}_{ge} \cdot
    \left ({\bf f}_{{\bf k},\eta}
    \hat a_{{\bf k},\eta}+ 
    {\bf f}^*_{{\bf k},\eta}
    \hat a^\dagger_{{\bf k},\eta}\right ),
\end{align}
where $\hat a^\dagger_{{\bf k},\eta}$ and $\hat a_{{\bf k},\eta}$
are the creation and annihilation operators for the fields
at wave vector ${\bf k}$
with polarization $\eta$.
The classical normal modes
can be written as
\begin{equation}
    {\bf f}_{{\bf k},\eta}
    = \frac{1}{\sqrt{V}}
    \hat \varepsilon_{{\bf k},\eta}e^{i {\bf k} \cdot {\bf r}},
\end{equation}
where $V$ is an arbitrary volume.

Beginning in the excited
state, $\ket{i}=\ket{e,\{{0}\}}$
and evolving to the final state
$\ket{f}=\ket{g,{\bf 1}_{{\bf k},\eta}}$,
the relevant matrix element for photon emission is
\begin{equation}
    \braket{e,\{{0}\}|H_{\rm int}|g,{\bf 1}_{{\bf k},\eta}}=
    \sqrt{\frac{\hbar\omega_{\bf k}}{2\epsilon_0 V}}
    \left (\hat \varepsilon_{{\bf k},\eta} \cdot
    {\bf d}_{ge} \right) e^{i{\bf k} \cdot {\bf r}}.
\end{equation}
Computing the free-space density of states
in the usual way, namely with periodic boundary conditions,
we have
\begin{equation}
D(\omega_0)
= \frac{\omega_0^2 V}{\pi^2 \hbar c^3}.
\end{equation}
Finally, using
$\omega_{\bf k} \approx\omega_0$
and $|\hat \varepsilon_{{\bf k},\eta} \cdot {\bf d}_{ge}|^2= |{ d}_{ge}|^2/3$ (isotropic averaging),
the SE rate is given by
\begin{equation}
    \gamma_0
    = \frac{\omega_0^3 |{\bf d}_{ge}|^2}{3 \pi\epsilon_0 \hbar c^3},
\end{equation}
which is identical to the 
$\gamma_0$ expressions in the main text (Eq.~\eqref{eq:SE0}),
and also 
with Eq.~\eqref{eq:qed_gamma_ab} when using the free-space Green's function.
Note in the quantum case, the dipole matrix {\it element}
is formally defined from 
 ${\bf d}_{ge} = \braket{g|{\bf \hat d}|e}$.
 
 \section{Electromagnetic fields generated by an oscillating electric dipole}
\label{APP_EM_dipole}
The exact equations that define the electric and magnetic fields generated by an idealized oscillating electric dipole located at the origin are given by
\begin{equation}\label{eq:analytical_E_dipole}
\begin{split}
\mathbf{E}(\mathbf{r}, t)=\frac{1}{4 \pi \epsilon_{0}}\Bigg[
&k^{2}(\hat{\mathbf{r}} \times \mathbf{d}) \times \hat{\mathbf{r}} \frac{e^{i k r}}{r} \\
&+[3(\hat{\mathbf{r}} \cdot \mathbf{d}) \hat{\mathbf{r}}-\mathbf{d}]\left(\frac{1}{r^{3}}-\frac{i k}{r^{2}}\right) e^{i k r}\Bigg]
\end{split}
\end{equation}
and
\begin{equation}
\mathbf{B}(\mathbf{r}, t)=\frac{\mu_{0}}{4 \pi}\Bigg[c k^{2}(\hat{\mathbf{r}} \times \mathbf{d}) \frac{e^{i k r}}{r}\left(1-\frac{1}{i k r}\right)\Bigg],
\end{equation}
where $k=\omega/c$ and $\omega$ is the angular frequency of the oscillating dipole, $\mathbf{d}=d_0 e^{-i\omega t}$ is the time-dependent dipole moment, ${r=|\mathbf{r}|}$, and $\hat{\mathbf{r}} = \mathbf{r}/r$~\cite{Jackson_1999}.

\bibliographystyle{elsarticle-num}
\bibliography{refs.bib}

\end{document}